\documentstyle[aps,epsf,rotate,multicol]{revtex}

\newcommand{\figdir}{./}
  
\begin{document}
\draft

\title{ Self-organized criticality in a rice-pile model}

\author{ Lu\'{\i}s~A.~Nunes Amaral$^{(1)}$ and Kent B{\ae}kgaard
Lauritsen$^{(2,3)}$}

\address{$^{(1)}$ Institut f\"ur Festk\"orperforschung,
                Forschungszentrum J\"ulich, D-52425 J\"ulich,
                Germany\\
$^{(2)}$ Center for Polymer Studies and Department of Physics, Boston
                University, Boston, Massachusetts 02215, USA\\
$^{(3)}$ Niels Bohr Institute, Center for Chaos and Turbulence
                Studies, Blegdamsvej 17, DK-2100 Copenhagen \O,
                Denmark}

\date{Submitted: February 8, 1996; Printed: \today}

\maketitle

\begin{abstract}
  We present a new model for relaxations in piles of granular material. 
The relaxations are determined by a stochastic rule which models the
effect of friction between the grains.
  We find power-law
  distributions for avalanche sizes and lifetimes characterized by the
  exponents $\tau = 1.53 \pm 0.05$ and $y = 1.84 \pm 0.05$,
  respectively.  For the discharge events, we find a characteristic
  size that scales with the system size as $L^\mu$, with $\mu = 1.20
  \pm 0.05$.  We also find that the frequency of the discharge events
  decrease with the system size as $L^{-\mu'}$ with $\mu' = 1.20 \pm
  0.05$.

\end{abstract}


\begin{multicols}{2}

  Since its introduction by Bak, Tang, and Wiesenfeld \cite{BTW}, the
concept of self-organized criticality (SOC) and models which display
SOC behavior have been the focus of much interest
\cite{BTW,KNWZ,CL,FF,OFC,Toner,Frette,HK,soc-rg}.  However, comparison
with real systems has proved to be a though test for the theory and
models \cite{JLN,H,RVK,RVR}.  Furthermore, in one dimension the models
tend to display either trivial behavior or behavior that cannot be
classified as critical.  Against this background, recent experiments
on rice piles \cite{Rice} have shown that under some conditions a real
rice pile can self-organize into a critical state:
For grains with a large aspect ratio the system self-organizes into a
critical state.  Frette {\it et al.\/} explained this result with the
increased friction and packing possibilities that were able to cancel
inertia effects.  Furthermore, they observed that 
large local slopes developed in the pile.

  Here, we propose a new model for a pile of granular material where 
we introduce randomness in the relaxation rule instead of in the
deposition rule.  We study the model in one dimension and find power
law distributions for avalanche sizes $s$ and lifetimes $T$.  We also
study the distribution of sizes for discharge events (i.e., particles
falling off the pile), and find it to be bounded.
The results show that our model belongs to a new universality class for
systems displaying SOC.

  First, we define the one-dimensional model. The system consists of a
plate of length $L$, with a wall at $i = 0$ and an open boundary at $i
= L+1$.  The profile of the pile evolves through two mechanisms:
deposition and relaxation.  Deposition is always done at $i = 1$, and
one grain at a time.  The rate of deposition is slow enough that any
avalanche, initiated by a deposited grain, will have ended before a
new grain is deposited.

  During relaxation we look at all {\it active\/} columns of the rice
pile: A column $i$ of the pile is considered active if, in the
anterior time step, it (i) received a grain from column $i-1$, (ii)
toppled a grain to column $i+1$, or (iii) column $i+1$ toppled one
grain to its right neighbor.  If a column $i$ is active {\it and\/}
the local slope, i.e., $\delta h(i) \equiv h(i) - h(i+1)$, is strictly
larger than a threshold value $S_1$, then with probability $p$ a grain
will move from $i$ to $i+1$.  However, if $\delta h(i) > S_2$, a grain
is moved from $i$ to $i+1$ with probability one.  Grains toppled from
column $i=L$ leave the system.  When no active columns remain on the
pile, the avalanche is said to be over.

\begin{figure}
\vspace*{-1.5cm}
\narrowtext
\centerline{
\epsfysize=1.1\columnwidth{\rotate[r]{\epsfbox{\figdir/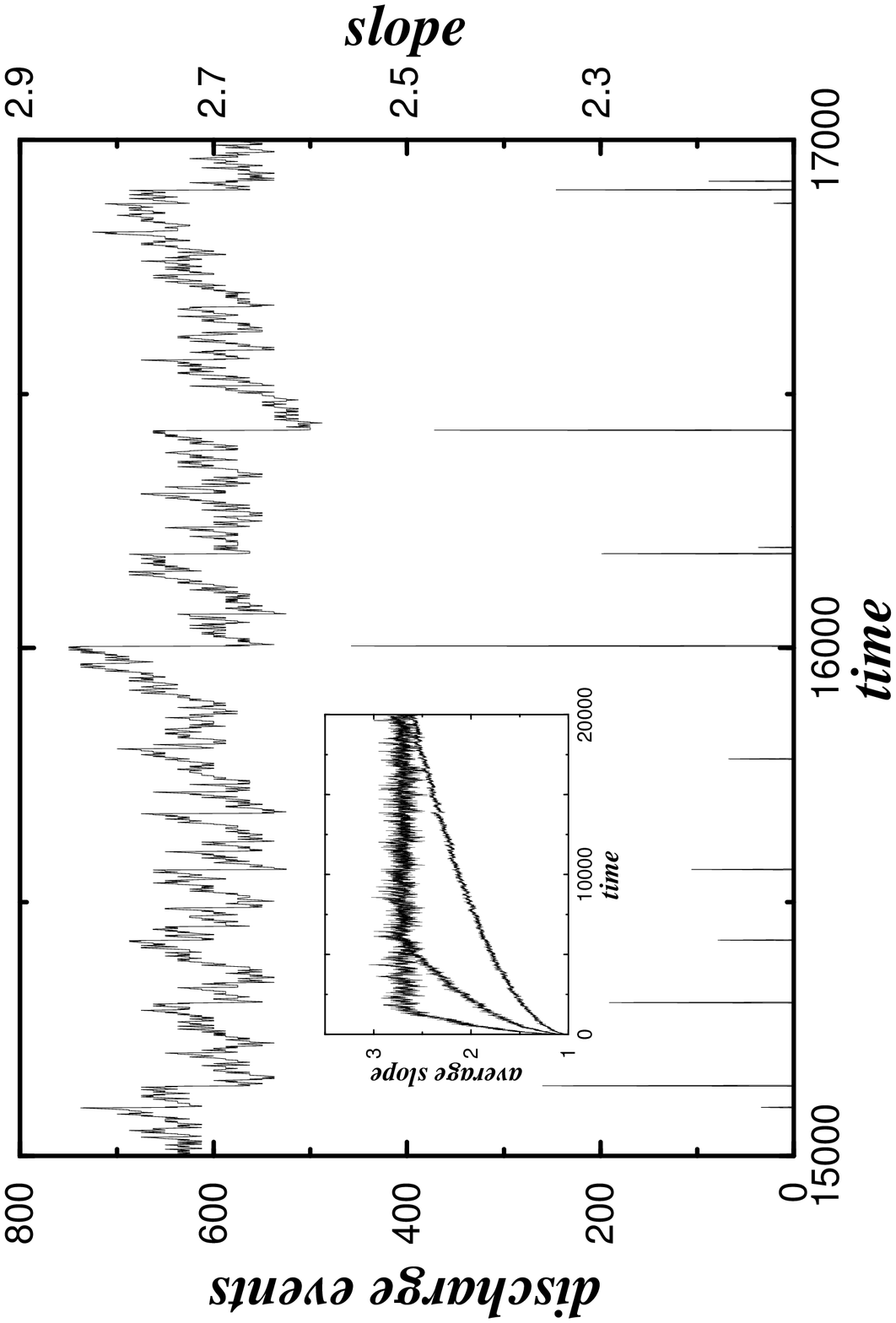}}}}
\centerline{
  \epsfysize=.8\columnwidth{\rotate[r]{\epsfbox{\figdir/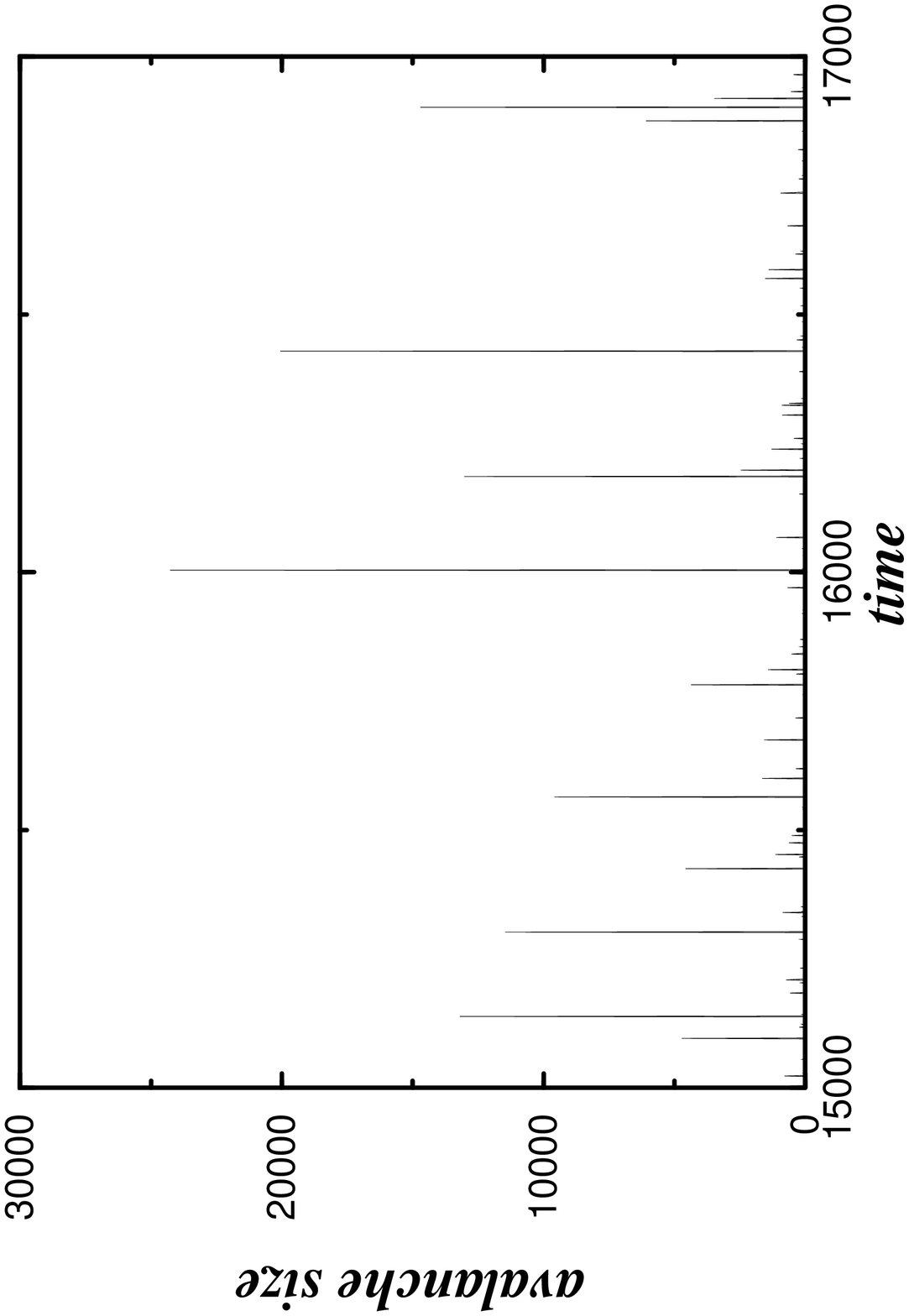}}}}
\vspace*{0.4cm}
\caption{ a) In the inset, we show the average slope as a function of
time (measured as the number of deposited grains).  As all the data
presented in this paper, the average slope was obtained for $p=0.6$,
$S_1=1$ and $S_2=4$.  The system sizes shown are $L=40, 80, 160$.  We
can see that after an initial transient regime, whose duration depends
strongly on the system size, a steady state is reached.  The figure
shows the fluctuations in the average slope of the pile, and the size
of the discharge events, as a function of time for as system of size
$80$.  It is visually apparent that there is a close connection
between sharp changes in the average slope and the discharge events at
the boundary.  It is interesting to note that although our model does
not include either a repose or a maximum angle for the pile, the
dynamics {\it suggest\/} the existence of such angles because of the
large discharge events.  b) Plot of the avalanches sizes as a function
of time for the same system and interval as in (a).  Again, the
connection between the largest avalanches and the discharge events is
observed.  }
\label{f-transient}
\end{figure}

  The physical interpretation of our rules is the following: 
Suppose that a column, or portion, of the pile is in a metastable
configuration.  If a new grain is deposited or toppled on top of it,
or the local slope changes, then that metastable configuration can
become unstable.  To model such an effect, we introduce the parameter
$p$, which represents the fact that there is a finite probability that
a new stable configuration is reached.  Physically the parameter $p$
thus describes the friction between the rice grains and the
possibility that a metastable packing configuration will be attained.
The friction effect is the main new ingredient in our model compared
to other models and it comes directly from the observation that there
exists a large range of slopes in the rice pile instead of a single
critical value \cite{Rice}.  The friction $p$ can be a complicated
function of local slopes and packings of the particles but we find
that the results are insensitive to the specific form and value of
$p$.  The parameter $S_2$ models the effect of gravity on the packing
arrangements.  We assume that above the maximum value $S_2$ of the
local slope, it is no longer possible for a local stable configuration
to be achieved, thus a grain must be toppled. In the limiting cases
$p=0, 1$, or $S_2=S_1$, we recover the model in Ref.\ \cite{BTW}
(which has trivial behavior for one dimension).

\begin{figure}
\narrowtext
\centerline{
  ~~\epsfysize=.8\columnwidth{\rotate[r]{\epsfbox{\figdir/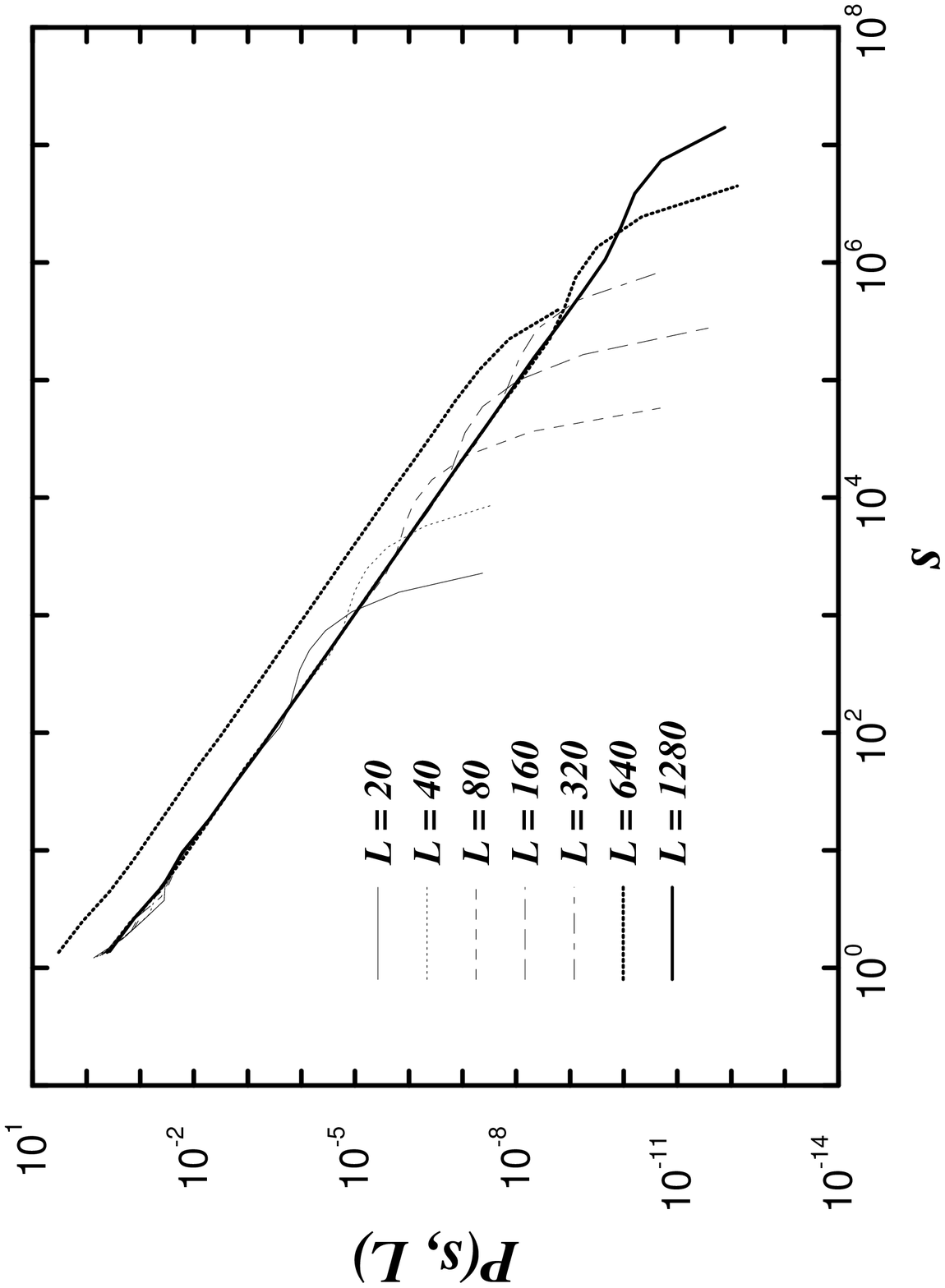}}}
}
\centerline{
  ~~\epsfysize=.8\columnwidth{\rotate[r]{\epsfbox{\figdir/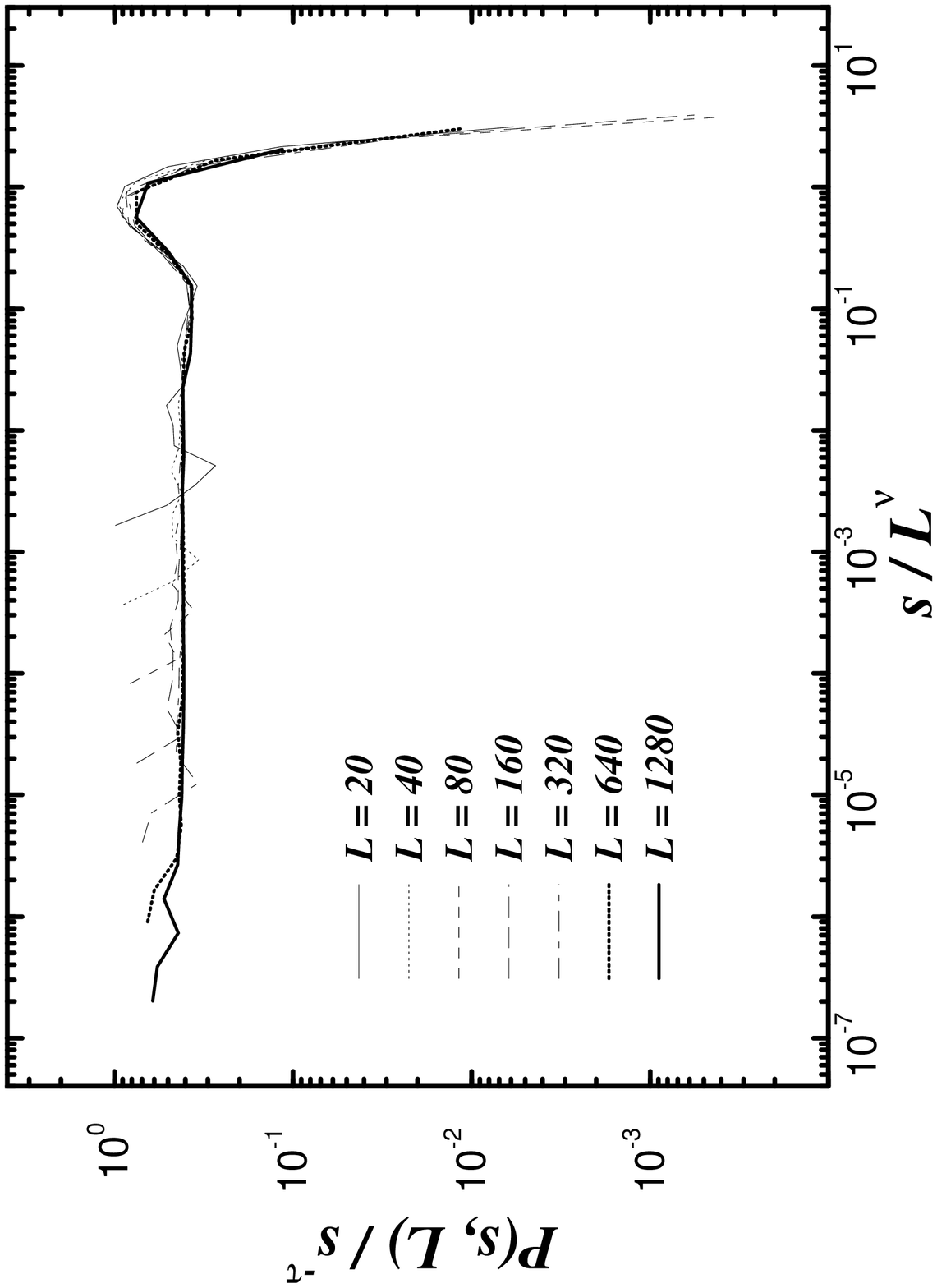}}}
}
\vspace*{0.4cm}
\caption{ a) Log-log plot of the probability density of avalanche
sizes $s$ for several system sizes.  It is visually apparent that for
$s \gg 1$ a power-law dependence is observed.  For values of $s$ close
to the cutoff, imposed by the finiteness of the system, we observe a
peak deviating from the power-law behavior.  To show that the peak is
due to the system reaching a supercritical state (which is followed by
a discharge) we also plot the distribution of avalanches when {\it
no\/} discharge event occurred: That curve, obtained for $L=640$, is
shifted vertically by a factor of $8$, to make it more visible.  b)
Data collapse of the curves shown in (a) according to
Eq.~(\protect\ref{e-internal}) with the exponents $\tau \simeq 1.53$
and $\nu \simeq 2.20$.  }
\label{f-internal}
\end{figure}

  The simulation of the model shows two distinct regimes, a transient
period followed by a steady (critical) state; cf.\
Fig.~\ref{f-transient}.  Here, we focus on the properties of the model
in the critical state.  As can be seen in Fig.~\ref{f-transient}, the
model leads to the establishment of a state with wildly varying
avalanches sizes and a complicated structure in time.  The size of an
avalanche can be defined in a number of ways: the number of topplings
$s$, the lifetime of the avalanche $T$, or the size of the discharge
events $m$.  We start by investigating the distribution of $s$.
Figure~\ref{f-internal}(a) shows the probability density of avalanche
sizes for different system sizes.  The distribution follows the
scaling form
\begin{equation}
        P(s,L) \sim s^{-\tau}~f_s(s / L^{\nu}),
						\label{e-internal}
\end{equation}
where $f_s$ is a scaling function rapidly decaying for large
arguments.  The best collapse is obtained with the exponents $\tau =
1.53 \pm 0.05$ and $\nu = 2.20 \pm 0.05$, cf.\
Fig.~\ref{f-internal}(b).  Even though $\tau$ is close to the
mean-field value 3/2 \cite{mf,alstr,ZLS}, our model describes a {\em
new\/} universality class.  This can be shown by mapping the avalanche
dynamics to the motion of an interface through a disordered medium
\cite{paczuski-boettcher:1996}.  By using that the average number of
topplings is $\left< s \right> = L$ in the critical state, it follows
from (\ref{e-internal}) that 
\begin{equation}
	\tau = 2 - \frac{1}{\nu} \, ,
\end{equation}
in agreement with our numerical results.

\begin{figure}
\narrowtext
\centerline{\epsfysize=.9\columnwidth{\rotate[r]{\epsfbox{\figdir/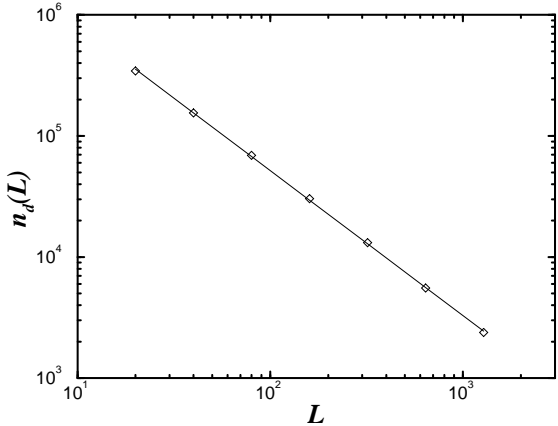}}}}
\vspace*{0.5cm}
\caption{The dependence of the number of discharge events $n_d$ as a
function of $L$.  A power law behavior, cf.\
Eq.~(\protect\ref{eq:n_d}), with $\mu' \simeq 1.20$ is obtained.  }
\label{f-discharge}
\end{figure}

  An interesting characteristic of the distribution is the presence of
a peak, deviating from the power-law behavior, for a size close to the
cutoff of the distribution.  A close look at Fig.~\ref{f-transient}
shows that the biggest avalanches coincide with large changes in the
average slope of the pile and with discharge events.  Furthermore, as
shown in Fig.~\ref{f-discharge}, the number $n_d$ of avalanches
reaching the open boundary, for a given number of deposited grains,
scales with the system size as
\begin{equation}
	n_d \sim L^{-\mu'} , 
				\label{eq:n_d}
\end{equation}
with $\mu' = 1.20 \pm 0.05$.  This suggests that the peak is due to
finite-size effects which lead the system into a supercritical state,
followed by a massive avalanche and a large change in the average
slope.  We check this hypothesis by considering {\it only\/}
the avalanches for which no discharge occurred at the boundary.  As
can be seen in Fig.~\ref{f-internal}(a), the peak for very large
values of $s$ is then no longer present and the cutoff has moved to a
smaller value, confirming our hypothesis.  We find that the data is
described by Eq.~(\ref{e-internal}) with the same values of
the exponents as when the peak is present.

\begin{figure}
\narrowtext
\centerline{\epsfysize=.9\columnwidth{\rotate[r]{\epsfbox{\figdir/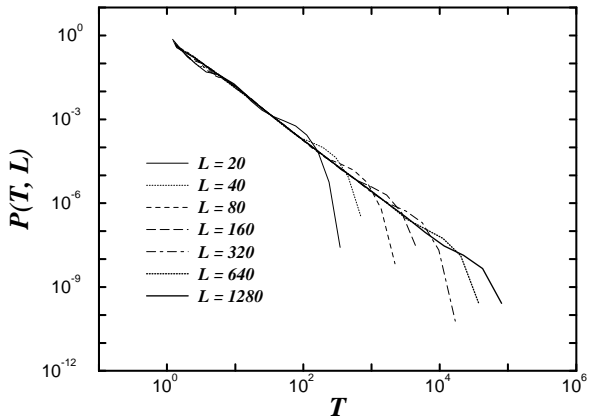}}}}
\centerline{\epsfysize=.9\columnwidth{\rotate[r]{\epsfbox{\figdir/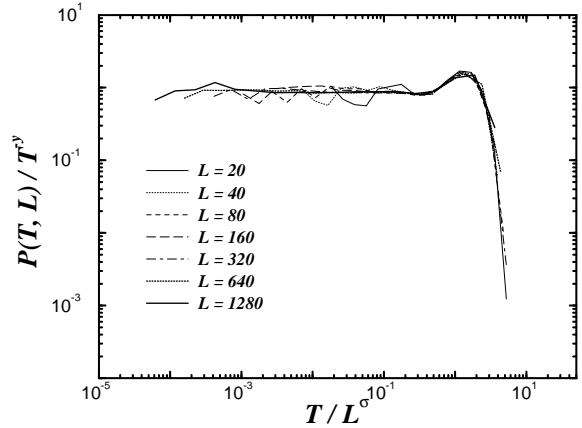}}}}
\vspace*{0.5cm}
\caption{ a) Log-log plot of the probability density of avalanche
lifetimes $T$ for several system sizes.  It is visually apparent that
for $t \gg 1$ a power-law dependence is observed.  As for the
avalanche sizes, a peak is present for lifetimes close to the cutoff.
b) Data collapse of the curves shown in (a) according to
Eq.~(\protect\ref{e-temporal}) with the exponents $y \simeq 1.84$ and
$\sigma \simeq 1.40$.  }
\label{f-temporal}
\end{figure}

\begin{figure}
\narrowtext
\centerline{\epsfysize=.9\columnwidth{\rotate[r]{\epsfbox{\figdir/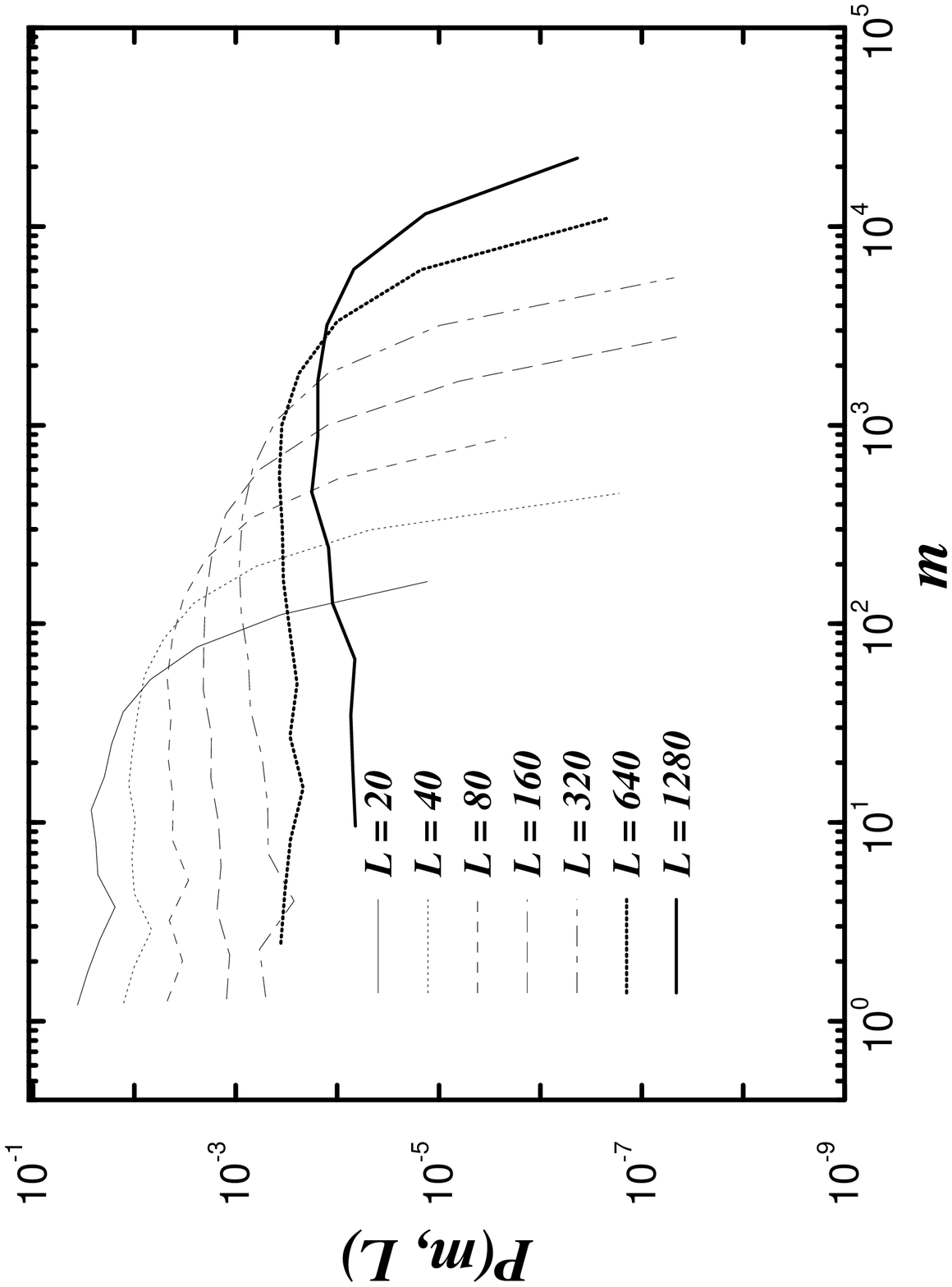}}}}
\centerline{\epsfysize=.9\columnwidth{\rotate[r]{\epsfbox{\figdir/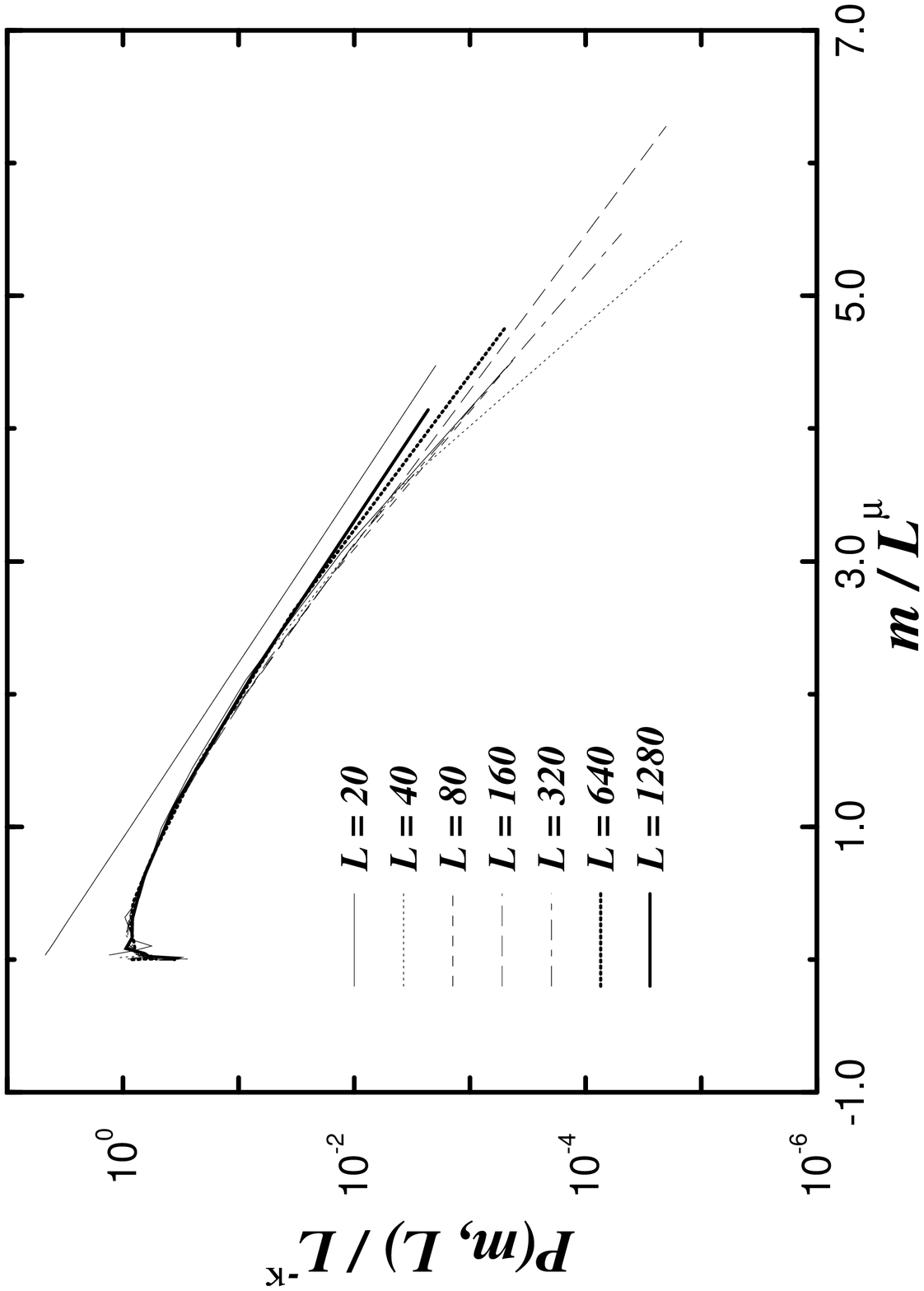}}}}
\vspace*{0.5cm}
\caption{ a) Log-log plot of the probability density of the sizes $m$
of the discharge events for several system sizes. It is visually
apparent that the distribution is bounded. b) Data collapse of the
curves shown in (a) according to Eq.~(\protect\ref{e-discharge}) with
the exponents $\kappa = \mu \simeq 1.20$.  As a visual aid, we display
a line corresponding to an exponential dependence.  }
\label{f-external}
\end{figure}

Next, we study the distribution of lifetimes $T$ for the avalanches.
As shown in Fig.~\ref{f-temporal}(a), the data is described by the
scaling form
\begin{equation}
          P(T, L) \sim T^{-y}~f_T(T / L^{\sigma}),
\label{e-temporal}
\end{equation}
which is confirmed by the good data collapse obtained with the
exponents $y = 1.84 \pm 0.05$ and $\sigma = 1.40 \pm 0.05$.  {}From
conservation of probability follows that $\sigma(y-1) = \nu(\tau-1)$
\cite{paczuski-boettcher:1996}, in nice agreement with our results.
Finally, we study the distribution of sizes $m$ for the discharge
events (see Fig.~\ref{f-external}(a)). The scaling ansatz
\begin{equation}
        P(m, L) \sim L^{-\kappa}~f_m(m / L^{\mu}),
\label{e-discharge}
\end{equation}
where the scaling function $f_m$ decays exponentially, describes the
data.  Since the distribution $P(m, L)$ does not diverge for $m \to 0$
and its integral must equal one, it follows that $\kappa = \mu$.  This
result is confirmed by the data collapse shown in
Fig.~\ref{f-external}(b), obtained for the exponents $\kappa = 1.2 \pm
0.1$ and $\mu = 1.2 \pm 0.1$.

  It is possible to obtain additional scaling relations for the
exponents besides those mentioned above.  In the steady state, the
input of matter must balance the output through the open boundary.
Thus, we obtain that the frequency of discharge events must balance
their characteristic size, and $\mu = \mu'$.  The characteristic size
of the discharge events depends on the system size as $L^{\mu}$.  So,
we can conclude that whenever the system reaches a supercritical
state, the number of grains discharged is of order $L^{\mu}$.  Since
the average number of topplings for a given grain before being
discharged is of order $L$, it follows that the cutoff size for the
avalanches must scale as $L \times L^{\mu} \sim L^\nu$, thus $\nu = 1
+ \mu$, in accordance with our results.

Just before sumbission of the present work, we became aware of a
model by Christensen et al.\ \cite{christensen-etal:1996}
which for some range of parameters
seems to belong to the same universality class as the model discussed
here.  The model in Ref.\ \cite{christensen-etal:1996}
introduces stochasticity in the
toppling of particles via the selection of a new random critical slope
for columns where a toppling occurred.  For local slopes above the
critical slope, a grain is always toppled.  
In Ref.\ \cite{christensen-etal:1996}
the predictions of the model are compared with
experimental results for the diffusion of tracer particles. 
The numerical value of the exponent describing the diffusion of the
tracers is in rough agreement with the numerical
predictions of the model. 
The exponent $\alpha$ describing the scaling of the potential
energy dissipated during an avalanche is $\alpha \simeq 1.53$
for our model \cite{AL},
in disagreement with the experimental result $\alpha \simeq 2$.  
In fact, the large
disagreement between the experimental value of $\alpha$ and the
numerical prediction of the model suggest that our model and the model
of Ref.\ \cite{christensen-etal:1996} do not belong to
the same universality class as the rice-pile experiment.
In \cite{luebeck-usadel}, a stochastic sandpile model is studied in
which only the front of the avalanches propagate (i.e., no backward
avalanches are allowed).  Such a rule leads to higher values for $\tau$
but apparently at the cost of destroying universality.

  Frette {\it et al.\/} also found that for ``round'' rice grains the
system did not evolve into a critical state, and that the distribution
of avalanche sizes was bounded.  The reason for this result can be
understood if the results for the role of inertia on the dynamics of
sandpiles are remembered \cite{KSG,PO}.  It was shown in
Ref. \cite{KSG} that the assumption of zero inertia is essential for
the establishment of the critical state.  For real experiments, where
inertia cannot be avoided, that assumption can only be valid for
system with sizes smaller than a threshold value $L_c$ \cite{KSG,PO}.
Thus, for the round rice grains all system sizes studied in the
experiments are larger than $L_c$, while for the elongated grains the
opposite is true.  Since our model has the implicit assumption of zero
inertia it is inevitable that we will only be able to investigate the
regime $L \ll L_c$, where SOC is observed.

  In summary, we present a new physically motivated model for piles of
granular material.  We find that the model self-organizes into a
critical state with distributions for most quantities described by
power laws.  We measure the exponents characterizing these
distributions, discuss scaling relations,
and find that our model belongs to a new universality class.

%Ack's

We acknowledge discussions with J. Krug, M. Mar\-ko\-sova, K. Sneppen,
H. E. Stanley, and S. Zapperi, and thank K. Christensen for
discussions concerning Ref.~\cite{Rice}.  K.B.L. thanks the Danish
Natural Science Research Council for financial support.

\end{multicols}

\end{document}